# Dynamics of a quantum emitter resonantly coupled to both external field and localized surface plasmon

Khachatur V. Nerkararyan[1], Torgom S. Yezekyan[2], and Sergey I. Bozhevolnyi[2,*]

[1]Department of Physics, Yerevan State University, 375049 Yerevan, Armenia

[2]Centre for Nano Optics, University of Southern Denmark, Campusvej 55, DK-5230 Odense M, Denmark

**Abstract**

We investigate excitation dynamics in the system of a quantum dipole emitter (QDE) coupled to a located nearby metal nanoparticle (MNP), which exhibits a dipolar localized surface plasmon (LSP) resonance at the frequency of the QDE radiative transition, in the presence of a strong external resonant electromagnetic field. Considering the QDE-field interactions in the regime of strong QDE-field coupling, we show that the feedback provided by the MNP on the QDE (due to the LSP excitation with the field generated by the dipole moment of the QDE transition) influences significantly the coherent process of Rabi oscillations, resulting in the occurrence of additional satellite frequencies in the radiation spectrum scattered by the QDE-MNP configuration. The relative ratio of high harmonics depends strongly on the QDE-MNP separation, an important characteristic feature that can be used for observing this effect and exploited, for example, for controlling distances at nanoscale.

# I. INTRODUCTION

Resonant interactions between quantum emitters placed in nanostructurted environment with strong electromagnetic fields bring about many interesting physical phenomena. Thus, it was found that electromagnetic coupling between quantum dipole emitters (QDEs), such as molecules, color centers in diamond or quantum dots, and surface plasmon polariton modes (often shortened to surface plasmons) allows control over the flow of electromagnetic energy [1-5]. QDE coupling to spatially localized surface plasmon excitations, localized surface plasmons (LSPs), supported by metal nanoparticles (MNPs) or, in general, nanostructures (that are often referred to as nanoantennas) results in strong modification of spontaneous emission rates in the regime of weak coupling [6-10], and can also lead to periodic coherent QDE-LSP energy transfer in the form of Rabi oscillations in the regime of strong coupling [11-13]. At the same time, electromagnetic interactions of atoms and molecules with intense laser fields that have been scrutinized during several decades [14] became again important in the context of coherent control of atom-photon interfaces that is vital in the realization of quantum information protocols [15].

Numerous investigations considering modification of spontaneous emission in various QDE-MNP configurations were focused on the effect of dramatic enhancements of the QDE (radiative and nonradiative) decay rates in the vicinity of the MNP [6-8, 16-20], always implicitly assuming that the relaxation dynamics is purely exponential as obtained in the Weisskopf-Wigner treatment of an individual two-level atom [21]. However, the LSP resonant excitation associated with free electron oscillations in an MNP can, for nm-size MNPs, be regarded as classical current oscillations, since a large number of free electrons ($\sim$100 nm$^{-3}$) are involved and their energy spectrum can be considered continuous. This classical oscillating current can then represented by a quantum coherent state of the LSP [22]. Note that the coherent LSP state is fundamentally different from (often-considered) LSP states with a definite number of quantized plasmons [23].

From the basic principles of quantum optics, it is known that quantized fields created by classical currents are described by a wave function of the coherent state [22]. These quantized fields are largely equivalent to classical fields, allowing one to employ the semi-classical approximation. Thus, in the studies of the relaxation dynamics of resonantly coupled QDE-MNP and QDE-MNP-QDE systems, it was found that the results obtained using both quantum [24, 25] and semiclassical [25, 26] approaches are identical when representing the LSP oscillating current by the coherent state in the quantum approach [24, 25]. Here, we employ the semiclassical [25-27] approach to investigate the excitation dynamics in the resonantly coupled QDE-MNP system illuminated with a strong external resonant electromagnetic field. Our analysis is based on the extension of coupled equations describing usual harmonic Rabi oscillations [22] so as to include an additional term responsible for the resonant QDE-MNP interaction [26]. The key feature of the resonant QDE-MNP coupling is that the particle polarizability becomes purely imaginary at resonance, so its response exhibits the $\pi/2$ phase shift. Considering the QDE-field interactions in the regime of strong coupling, we show that the feedback provided by the MNP on the QDE (due to the LSP excitation with the field generated by the dipole moment of the QDE transition) influences significantly the coherent process of Rabi oscillations, resulting in the occurrence of additional satellite frequencies in the radiation spectrum scattered by the QDE-MNP configuration.

## II. THEORETICAL FRAMEWORK

The system under consideration is schematically presented in Fig. 1 and consists of a two-level QDE and a spherical MNP, which interacts resonantly with the QDE and external electromagnetic field. It should be noted that, at the initial moment of switching on the external field, the QDE can be found both in the ground and excited states. In addition, we assume that the QDE-MNP coupling and the external pump field are sufficiently strong so that the relaxation processes can be neglected, at least at the initial stage. It is further assumed that the spherical MNP exhibits a dipolar LSP resonance at the frequency $\omega$ of the radiative (dipole-allowed) transition between the excited and ground states. In the presence of the pump field, the QDE wave function can be represented as the coherent superposition state:

$$\Psi(t) = a_1(t)\phi_1 \exp\left(-\frac{i}{\hbar}E_1 t\right) + a_0(t)\phi_0 \exp\left(-\frac{i}{\hbar}E_0 t\right) , \qquad (1)$$

where $\phi_1$ and $\phi_0$ are the wave functions of the QDE in the excited and ground states characterized by the energies $E_1$ and $E_0$, respectively, while $a_1(t)$ and $a_0(t)$ are the corresponding (time-dependent) probability amplitudes. The QDE transition dipole moment is thereby given by

$$\vec{D} = a_1 a_0^* \vec{d}_{10} \exp(-i\omega t) + a_0 a_1^* \vec{d}_{10}^* \exp(i\omega t) , \qquad (2)$$

with the asterisk denoting the complex conjugate, and $\vec{d}_{10} = \int \phi_1 e\vec{r} \phi_0^* dV$ and $\hbar\omega = E_1 - E_0$ being the dipole moment and energy of the transition between the excited and ground states, respectively. The external resonant pump field can be represented as follows:

$$\vec{E}_{ext} = \vec{E}_0 \exp(-i\omega t) + \vec{E}_0^* \exp(i\omega t) , \qquad (3)$$

We assume that the radiation wavelength $\lambda$ is significantly larger than the distance $R$ between the QDE and MNP center, which is in turn much larger than the MNP radius $r$ (Fig. 1(a)]: $\lambda \gg R \gg r$. In this electrostatic approximation, the MNP can be considered as being subjected to the homogeneous electric field $\vec{E}_{tot}$ created by the incident external and QDE scattered fields. For the QDE transition dipole moment $\vec{D}$ [Eq. (2)] oriented along the QDE-MNP symmetry axis, this (total) field can be expressed in the following form:

$$\vec{E}_{tot} = \vec{E}_{ext} + \frac{\vec{D}}{2\pi\varepsilon_0\varepsilon_2 R^3} \ . \qquad (4)$$

Here, $\varepsilon_0$ and $\varepsilon_2$ are the relative permittivities of vacuum and the dielectric environment. In general, the MNP response should be determined by considering the corresponding dynamics influenced by the total external field and the LSP relaxation. However, as noted in our previous work [24-27], the MNP response can be considered instantaneous due to extremely fast relaxation of the LSP excitation, since the LSP lifetime is on the fs-scale while the QDE lifetime (even shortened due to the MNP proximity) would be on the ns-scale [6-8]. The total field acting on the QDE is a sum of the external pump field, the MNP scattered external field and the feedback QDE field produced by MNP scattering back to the QDE:

$$\vec{E} = \frac{2(\varepsilon_1 - \varepsilon_2)}{(\varepsilon_1 + 2\varepsilon_2)} \frac{r^3}{R^3} \left( \vec{E}_0 e^{-i\omega t} + \frac{1}{2\pi\varepsilon_0\varepsilon_2 R^3} a_1 a_0^* \vec{d}_{10} e^{-i\omega t} \right) + \vec{E}_0 e^{-i\omega t} + \text{c.c.} \ , \qquad (5)$$

where $\varepsilon_1 = \varepsilon_{1r} + i\varepsilon_{1i}$ is the MNP relative permittivity, and c.c. stands for complex conjugate.

Using the time-dependent Schrödinger equation for two-level systems in the driving field given by Eq. (5) and carrying out standard manipulations within the rotating wave approximation, one obtains the following system of coupled equations for the probability amplitudes ($\dot{a} \equiv da/dt$):

$$\dot{a}_0 = \mu a_1^* a_1 a_0 + \beta a_1 \ , \qquad (6a)$$

$$\dot{a}_1 = -\mu a_0^* a_0 a_1 - \beta^* a_0 \ , \qquad (6b)$$

where $\mu$ characterizes the feedback produced by the MNP scattering back to the QDE:

$$\mu = \frac{3}{\pi\hbar\varepsilon_0\varepsilon_{1i}} \frac{r^3}{R^6} \left|\vec{d}_{10}\right|^2 \ , \qquad (7)$$

and $\beta$ characterizes the strength of the external resonant pump field acting together with the MNP response on the QDE:

$$\beta = \left[ \frac{6\varepsilon_2}{\hbar \varepsilon_{1i}} \cdot \frac{r^3}{R^3} - \frac{i}{\hbar} \right] \vec{d}_{10} \cdot \vec{E}_0^* \ . \tag{8}$$

In obtaining the above formulae, we assumed in Eq. (5) resonant QDE-MNP interaction, implying that $\mathrm{Re}(\varepsilon_1 + 2\varepsilon_2) = 0$, and relatively low LSP damping, implying that $3\varepsilon_2 \gg \varepsilon_{1i}$ [24-26]. Here we introduced rate parameters $\mu$ and $\beta$ that allow us to simplify the coupled rate equations above [Eqs. (6a) and (6b)], characterizing the influence of QDE coupling to the MNP and pump field, respectively, on the QDE dynamics. The QDE-MNP coupling decreases for larger QDE-MNP separations and weaker QDE dipole moments [Eq. (7)], and so does parameter $\mu$: $\mu \to 0$ when $R \to \infty$ and/or $|\vec{d}_{10}| \to 0$. The QDE-field coupling decreases for weaker fields and QDE dipole moments [Eq. (8)], and so does parameter $\beta$: $\beta \to 0$ when $E_0 \to 0$ and/or $|\vec{d}_{10}| \to 0$. Note that the QDE-field coupling involves also the pump field scattered by the MNP towards the QDE [see the first term in brackets of Eq. (8)], so that one cannot adjust independently these two rate parameters. Generally speaking, parameters $\mu$ and $\beta$ represent Rabi frequencies for the optical fields (acting on the QDE) originating from the MNP feedback and driven by the external pump field, respectively. Combining Eqs. (6a) and (6b) results in

$$a_1 \dot{a}_0 - a_0 \dot{a}_1 = \mu a_1 a_0 + \beta a_1^2 + \beta^* a_0^2 \ , \tag{9}$$

along with $|a_0|^2 + |a_1|^2 = \mathrm{const}$. This constant is to be set to one, bearing in mind that $a_1$ and $a_0$ are the probability amplitudes.

One of the most important assumptions made is related to the strength of the QDE-MNP coupling which should ensure considerably larger relaxation rates than that for the QDE in free space. The corresponding ratio can be evaluated now with the help of Eq. (7) and the Weisskopf-Wigner result [21] as follows:

$$\eta = \frac{\mu}{\gamma_0} = \frac{9}{\varepsilon_{1i} \sqrt{\varepsilon_2}} \left( \frac{\lambda}{2\pi R} \right)^3 \left( \frac{r}{R} \right)^3 \ , \tag{10}$$

where $\gamma_0$ is the QDE spontaneous decay rate in vacuum [21], and $\lambda$ is the vacuum wavelength corresponding to the QDE transition frequency $\omega$. For a typical dielectric environment with $\varepsilon_2 = 2.25$ (e.g., glass or polymer), the resonance condition (i.e., $\varepsilon_{1r} = -4.5$) is met, for gold, at the wavelength of ~ 530 nm with $\varepsilon_{1i}^g \cong 2.35$ and, for silver, at ~ 400 nm with $\varepsilon_{1i}^s \cong 0.22$ according to the experimental data presented in [28]. Considering an MNP with the radius of 5 nm and the QDE distance to the MNP center being 15 nm (in order to be within the electrostatic dipole description), one obtains the ratio $\eta \approx 17$ for gold and $\eta \approx 77$ for silver, justifying thereby the aforementioned assumption: $\mu \gg \gamma_0$. It is interesting that the effect is already pronounced at relatively large (~ 10 nm) distances between QDEs and the MNP surface, which are in the range of distances explored in the recent experiments with 10-nm-size gold nanoparticles [17]. It is also transparent that even larger ratios can be achieved by exploiting the LSP shape dependence [11] and red-shifting the MNP resonance towards smaller metal absorption [28].

### III. RESULTS AND DISCUSSION

Analyzing coupled equations [Eqs. (6a) and (6b)], one realizes that their structure becomes radically different in two extreme cases: $\mu = 0$ and $\beta = 0$. In the first case, one is left with the equations describing usual (harmonic) Rabi oscillations [22], whereas the second case corresponds exactly to the situation that we considered previously [26] – the resonant QDE-MNP interaction (in the absence of the pump field) resulting in the dynamics following a non-oscillatory (and non-exponential) decay. One should expect therefore that a change from oscillatory to non-oscillatory behavior occurs at a certain ratio between these rate parameters, which represents, from the physical viewpoint, a ratio between the strength of the external resonant pump field acting (together with the MNP response) on the QDE and the feedback produced by the MNP scattering back to the QDE. Selecting this ratio with the aim of establishing a clear demarcation line between these two regimes, one arrives at the following parameter: $\alpha = 2|\beta|/\mu$, which becomes thereby the most important parameter of the considered configuration. This parameter determines whether the configuration dynamics follows the regime of weak ($\alpha < 1$) or strong ($\alpha > 1$) pump. For weak pump fields, the time evolution of the probability to find the QDE in the excited state can be expressed as follows:

$$|a_1(t)|^2 = \frac{\alpha^2}{\alpha^2 + \left[1 + \sqrt{1-\alpha^2} \cdot \coth(0.5\delta\tau - \sigma)\right]^2} , \qquad (11)$$

where

$$\delta = \eta\sqrt{1-\alpha^2}, \text{ and } \tau = \gamma_0 t. \qquad (12)$$

Here, $\sigma = \tanh^{-1}\left(\sqrt{1-\alpha^2}\right)$, when $|a_1(0)|=1$, and $\sigma = 0$, when $|a_1(0)|=0$. It is seen that the probability of finding the QDE in the excited state approaches in the long run a stationary value:

$$|a_1(t)|^2 \to \frac{\alpha^2}{2\left(1+\sqrt{1-\alpha^2}\right)} \text{ when } t \to \infty. \qquad (13)$$

Here, one should bear in mind that the time scale considered has to be kept below the spontaneous emission time, i.e., one has to maintain the following inequality: $\tau = \gamma_0 t \ll 1$. For longer times, the usual (Weisskopf-Wigner) mechanism of spontaneous emission [21] can no longer be ignored. Overall, in the case of weak pump ($\alpha < 1$), regardless of the initial conditions the system evolves towards a stationary superposition state (Fig. 2), with the excited state probability increasing from 0 to 0.5 for the values of $\alpha$ increasing from 0 to 1 [Eq.(13)]. This rather remarkable feature is associated with the role played by the feedback produced by the MNP scattering back to the QDE, which promotes the QDE transition from the excited to the ground state and prevents a reverse transition. As already pointed out in the beginning of this section, small values of $\alpha$ (due to the dominance of the MNP scattering over weak pump fields) result in the dynamics very similar to that predicted for the resonant QDE-MNP interaction in the absence of the pump field [26], which is characterized by non-oscillatory decay towards the ground state (also independently on the initial conditions). The non-oscillatory behavior in this case, as well as in other similar systems without external fields, is caused by the $\pi/2$ phase delay in the resonant response of the MNP on the acting (external and scattered by the QDE) fields [24-26].

In the case of strong pump fields ($\alpha > 1$), the QDE dynamics changes drastically becoming governed by Rabi oscillations, i.e., the strongly pumped QDE switches periodically between the excited and ground states. For the time dependence of the probability of finding the QDE in the excited state, one obtains correspondingly:

$$|a_1(t)|^2 = \frac{\alpha^2}{\alpha^2 + \left[1+\sqrt{\alpha^2-1}\cot(0.5\Omega\tau-\varphi)\right]^2}, \qquad (14)$$

where

$$\Omega = \eta\sqrt{\alpha^2 - 1} \ . \tag{15}$$

Here, $\varphi = \tan^{-1}\left(\sqrt{\alpha^2 - 1}\right)$, when $|a_1(0)| = 1$, and $\varphi = 0$, when $|a_1(0)| = 0$. It is seen that the QDE dynamics becomes purely harmonic, i.e., exhibiting Rabi oscillations [22], in the limit of very strong pump fields, when $\alpha \to \infty$:

$$|a_1(t)|^2 = \sin^2(0.5\Omega\tau - \varphi) \ . \tag{16}$$

One can straightforwardly recover that the oscillation frequency in this limit and for weak MNP feedback, i.e., when $6\varepsilon_2 r^3 \ll \varepsilon_{1i} R^3$ [Eq. (8)], is given by the well-known formula for the Rabi oscillations of a two-level atom in a strong resonant electromagnetic field [22]. This change from the non-oscillatory behavior to Rabi oscillations caused by turning on the pump power is somewhat similar to that (also described by a semiclassical approach) occurred in the system of a photonic cavity resonantly coupled to a QDE and caused by increasing the cavity quality factor [29]. Both transitions are related to changes in the balance between the strengths of dissipation and resonant (pump or vacuum cavity) fields.

In the general case (Eq. [14]), however, the feedback provided by the MNP on the QDE (due to the LSP excitation by the QDE dipole moment) influences significantly the coherent process of Rabi oscillations, resulting in anharmonicity that is especially pronounced for values of $\alpha$ being close to 1 (Fig.3). Again, the $\pi/2$ phase delay in the resonant response of the MNP on the acting (external and scattered by the QDE) fields slows down the QDE excitation, while facilitating the QDE relaxation back to the ground state [24-26]. In this case, the initial conditions influence only the phase of Rabi oscillations, shifting the periodic response correspondingly along the time axis. The anharmonic behavior in the QDE dynamics results in an anharmonic oscillations of the QDE dipole moment, which determines the spontaneous emission characteristics and whose main component can be represented as follows:

$$p \sim -\frac{\beta}{|\beta|} a_1 a_0^* = b(\tau) = \frac{\alpha\left(1 + \sqrt{\alpha^2 - 1}\cot(0.5\Omega\tau - \varphi)\right)}{\alpha^2 + \left[1 + \sqrt{\alpha^2 - 1}\cot(0.5\Omega\tau - \varphi)\right]^2} \ . \tag{17}$$

It is seen (Fig. 4) that the dipole moment amplitude oscillates as expected for the case of strong coherent resonant pump, featuring anharmonicity that is directly connected with the aforementioned anharmonicity in the population dynamics. Also in this case, the initial conditions determine only the phase of oscillations.

The anharmonicity in the QDE dynamics, being dependent on $\alpha = 2|\beta|/\mu$ and especially pronounced for the values close to 1 (Figs. 3 and 4), depends thereby strongly on the QDE-MNP separation $R$ [see Eq. (8)]. This is an important characteristic feature that can be used for observing this effect and exploited, for example, for controlling distances at nanoscale [30]. Considering the QDE emission spectrum, it is clear that high harmonics would appear due to the anharmonicity at the multiple Rabi-shifted frequencies:

$$\omega_n = \omega \pm n\Omega_R, \quad \Omega_R = \Omega\gamma_0, \quad n = 0,1,2,\ldots. \tag{18}$$

The amplitudes of these harmonics can be chosen as the experimentally observable characteristics. These amplitudes can straightforwardly be found by expanding the dipole moment expression from Eq. (17) in a Fourier series:

$$b(\tau) = A_0 + \sum_{n=1}^{\infty} A_n \sin(n\Omega\tau + \vartheta_n). \tag{19}$$

From previous considerations, it is expected that the amplitude $A_n$ would most strongly depend on parameter $\alpha$ for its values close to 1 as indeed seen in Fig. 5. Also as expected, high harmonics decrease in amplitudes when increasing the external pump field and correspondingly increasing parameter $\alpha$ (Fig. 5).

Experimental observation of high harmonics requires that the Rabi frequency $\Omega_R$ [Eq. (18)] exceeds significantly the QDE relaxation rate $\gamma$ in the presence of MNP: $\Omega_R \gg \gamma$. It should be borne in mind that, in the considered configuration, the frequency of the QDE transition coincides with the LSP resonance, so that the QDE relaxation rate in the weak-coupling regime reads [31]:

$$\gamma = \left[1 + \left(\frac{r^3}{R^3} \cdot \frac{6\varepsilon_2}{\varepsilon_{1i}}\right)^2\right]\gamma_0. \tag{20}$$

Using the same configuration parameters as above one finds that, for a 5-nm-radius gold MNP, $\Omega_R \approx 28\gamma$ when $\alpha = 2$. Considering the transition dipole moment $d_{10} = 4.8 \cdot 10^{-29}\,\text{C}\cdot\text{m}$ (ensuring the spontaneous emission rate of $10^9\,\text{s}^{-1}$ at the 400-nm wavelength), one finds that the latter value of $\alpha$ is in turn achieved at laser fields with the amplitude $|\vec{E}_0| \approx 2.5 \cdot 10^4\,\text{V/m}$, implying the intensity of $\sim 0.5 \cdot 10^3\,\text{W/cm}^2$, i.e., a relatively low intensity easily achievable in laboratory conditions. It is thereby seen that the above condition for the observation of high harmonics is feasible to satisfy. Furthermore, given strong dependences of the relative ratios of high harmonics on parameter $\alpha$ (e.g., the ratio $A_1/A_3$ changes from 1.7 to 14.6 in the range shown in Fig. 5), one becomes interested in assessing the possibility of using this characteristic feature for monitoring distances at nanoscale. As the first step in this direction, one should directly relate variations in parameter $\alpha$ to changes in the QDE-MNP distance. Since the corresponding analytic expression is cumbersome:

$$\alpha(R) = \frac{\pi \varepsilon_2 \varepsilon_{1i}}{3} \cdot \frac{R^6}{r^3} \cdot \frac{E_0}{d} \sqrt{1 + \left(\frac{6\varepsilon_2}{\varepsilon_{1i}} \cdot \frac{r^3}{R^3}\right)^2} \,, \qquad (21)$$

we calculated a few dependencies using the same configuration parameters as above targeting the (most interesting) range of relatively small values of $\alpha$ (Fig. 6). It is readily seen that changing the distance by a couple of nm can produce the whole range of $\alpha$ considered in Fig. 5, indicating thereby a very high sensitivity of the relative ratios of high harmonics in Rabi oscillations with respect to the QDE-MNP distance. In practice, controlled approach of a QDE to an MNP can be realized by following, for example, the early concept of an MNP on a fiber tip [7, 8] or recent developments in plasmonic nano-optical tweezers [32].

## IV. CONCLUSIONS

We have reported a semiclassical consideration of excitation dynamics in the system of a QDE coupled to a located nearby MNP, which exhibits a dipolar LSP resonance at the frequency of the QDE radiative transition, in the presence of a strong external resonant electromagnetic field. Depending on the ratio $\alpha$ between the strength of the external resonant pump field acting (together with the MNP response) on the QDE and the feedback produced by the MNP scattering back to the QDE, the QDE dynamics follows either the weak pump regime characterized by a non-oscillatory dynamics with the QDE quickly reaching a stationary superposition state (in which the QDE excited state probability can become close to 0.5) or the strong pump regime characterized by Rabi oscillations with the pronounced anharmonicity. The latter regime results in the QDE scattering at additional frequencies (multiple of the Rabi frequency) with the relative ratio of high harmonics depending strongly on parameter $\alpha$ and thereby on the QDE-MNP separation, an important characteristic feature that can be used for observing this effect and exploited, for example, for controlling distances at nanoscale as discussed above. Another interesting possibility would be to exploit the transition between the two regimes by simply moving an MNP or a QDE with respect to each other for realization of quantum opto-mechanical effects capitalizing on this threshold behavior, or for realization of quantum logical operations. Overall, we believe that the reported results have far reaching implications within the very rapidly developing field of quantum plasmonics [23, 33].

## ACKNOWLEDGMENTS

The authors gratefully acknowledge financial support for this work from the European Research Council, Grant No. 341054 (PLAQNAP), as well as partial support (KVN) from the from the University of Southern Denmark (project SDU 2020).

**References**

[1]. D. E. Chang, A. S. Sørensen, P. R. Hemmer, and M. D. Lukin, Phys. Rev. Lett. **97**, 053002 (2006).

[2]. A.V. Akimov, A. Mukherjee, C. L. Yu, D. E. Chang, A. S. Zibrov, P. R. Hemmer, H. Park, and M. D. Lukin, Nature **450**, 402 (2007).

[3]. R. Kolesov, B. Grotz, G. Balasubramanian, R. J. Stöhr, A. A. L. Nicolet, P. R. Hemmer, F. Jelezko, and J. Wrachtrup, Nature Physics **5**, 470 (2009).

[4] A. Sipahigil, R. E. Evans, D. D. Sukachev, M. J. Burek, J. Borregaard, M. K. Bhaskar, C. T. Nguyen, J. L. Pacheco, H. A. Atikian, C. Meuwly, R. M. Camacho, F. Jelezko, E. Bielejec, H. Park, M. Lončar, and M. D. Lukin, Science **354**, 847 (2016).

[5] S. Kumar, V. A. Davydov, V. N. Agafonov, and S. I. Bozhevolnyi, Opt. Mater. Express **7**, 2586 (2017).

[6] J. N. Farahani, D. W. Pohl, H. J. Eisler, and B. Hecht, Phys. Rev. Lett. **95**, 017402 (2005).

[7] P. Anger, P. Bharadwaj, and L. Novotny, Phys. Rev. Lett. **96**, 113002 (2006).

[8] S. Kühn, U. Håkanson, L. Rogobete, and V. Sandoghdar, Phys. Rev. Lett. **97**, 017402 (2006).

[9] T. B. Hoang, G. M. Akselrod, and M. H. Mikkelsen, Nano Lett. **16**, 270 (2016).

[10] S. K. H. Andersen, S. Kumar, and S. I. Bozhevolnyi, Nano Lett. **17**, 3889 (2017).

[11] A. Trügler and U. Hohenester, Phys. Rev. B **77**, 115403 (2008).

[12] R. Chikkaraddy, B. de Nijs, F. Benz, S. J. Barrow, O. A. Scherman, E. Rosta, A. Demetriadou, P. Fox, O. Hess, and J. J. Baumberg, Nature **535**, 127 (2016).

[13] P. Törmä and W. L. Barnes, Rep. Prog. Phys. **78**, 013901 (2015), and references therein.

[14] K. Burnett, V. C. Reed, and P. L. Knight, J. Phys. B: At. Mol. Opt. Phys. **26**, 561 (1993).

[15] Y. Zhou, A. Rasmita, K. Li, Q. Xiong, I. Aharonovich, Nat. Commun. **8**, 14451 (2017).

[16] X. W. Chen, M. Agio and V. Sandoghdar, Phys. Rev. Lett. **108**, 233001 (2012).

[17]  G. P. Acuna, M. Bucher, I. H. Stein, C. Steinhauer, A. Kuzyk, P. Holzmeister,  R. Schreiber, A. Moroz, F. D. Stefani, T. Liedl, F. C. Simmel and P. Tinnefeld, ACS Nano **6**, 3189 (2012).

[18] K. E. Dorfman, P. K. Jha, D. V. Voronine, P. Genevet, F. Capasso and M. O. Scully, Phys. Rev. Lett. **111**, 043601 (2013).

[19] C. Sauvan, J. P. Hugonin, I. S. Maksymov and P. Lalanne, Phys. Rev. Lett. **110**, 237401 (2013).

[20] S. D'Agostino, F. D. Sala and L. C. Andreani, Phys. Rev. B **87**, 205413 (2013).

[21] V. F. Weisskof and E. P. Wigner, Z. Phys. **63**, 54 (1930).

[22] C. Gerry and P. Knight, *Introductory Quantum Optics* (Cambridge University Press, Cambridge, 2005).

[23] M. S. Tame, K. R. McEnery, S̨. K. Özdemir, J. Lee, S. A. Maier and M. S. Kim, Nat. Phys. **9**, 329 (2013).

[24] K. V. Nerkararyan and S. I. Bozhevolnyi, Faraday Discuss. **178**, 295 (2015).

[25] K. V. Nerkararyan and S. I. Bozhevolnyi, Phys. Rev B **92**, 045410 (2015).

[26] K. V. Nerkararyan and S. I. Bozhevolnyi, Opt. Lett. **39**, 1617 (2014).

[27] K. V. Nerkararyan, T. S. Yezekyan, and S. I. Bozhevolnyi, J. Luminescence **192**, 595 (2017).

[28] P. B. Johnson and R. W. Christy, Phys. Rev. B **6**, 4370 (1972).

[29] S. I. Bozhevolnyi and J. B. Khurgin, Optica **3**, 1418 (2016).

[30] J. Seelig, K. Leslie, A. Renn, S. Kühn, V. Jacobsen, M. van de Corput, C. Wyman, and V. Sandoghdar, Nano Lett. **7**, 685 (2007).

[31] P. Bharadwaj and L. Novotny, Opt. Express **15**, 14266 (2007).

[32] M. L. Juan, M. Righini, and R. Quidant, Nat. Photon. **5**, 349 (2011).

[33] S. I. Bozhevolnyi and J. B. Khurgin, Nat. Photon. **11**, 398 (2017).

**Figure captions**

FIG. 1. (Color online) Schematic of a system consisting of a QDEs placed near an MNP illuminated by a resonant external pump field, indicating (a) system parameters and (b) QDE energetic levels along with an oscillating current associated with the LSP excitation.

FIG. 2. (Color online) The probability of a QDE to be found in the excited state as a function of the normalized time in the weak pump regime for different values of parameter $\alpha < 1$ and initial conditions.

FIG. 3. (Color online) The probability of a QDE to be found in the excited state as a function of the normalized time in the strong pump regime for different values of parameter $\alpha > 1$ and initial conditions.

FIG. 4. (Color online) The dipole moment amplitude as a function of the normalized time in the strong pump regime for different values of parameter $\alpha > 1$ and initial conditions.

FIG. 5. (Color online) Dipole amplitude harmonics due to the anharmonicity in Rabi oscillations as a function of parameter $\alpha > 1$.

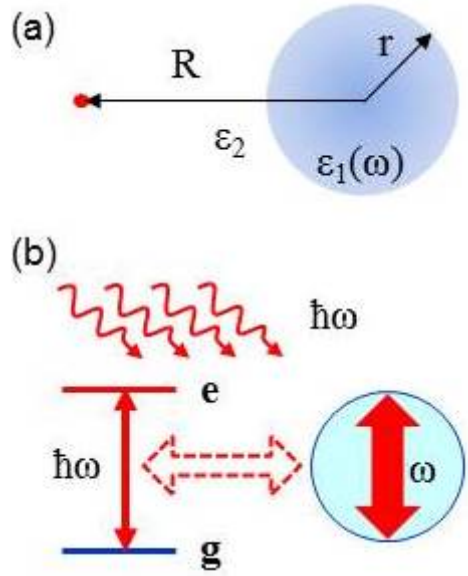

FIG. 1. (Color online) Schematic of a system consisting of a QDEs placed near an MNP illuminated by a resonant external pump field, indicating (a) system parameters and (b) QDE energetic levels along with an oscillating current associated with the LSP excitation.

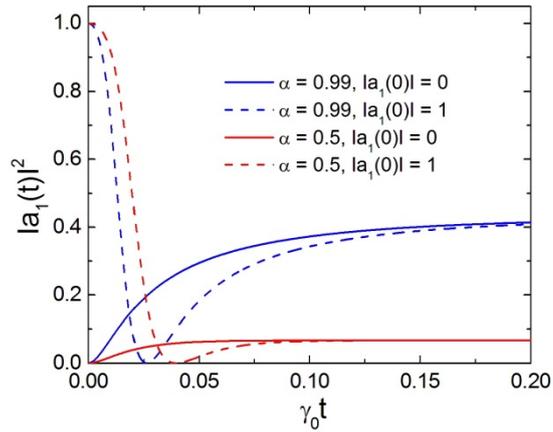

FIG. 2. (Color online) The probability of a QDE to be found in the excited state as a function of the normalized time in the weak pump regime for different values of parameter $\alpha < 1$ and initial conditions.

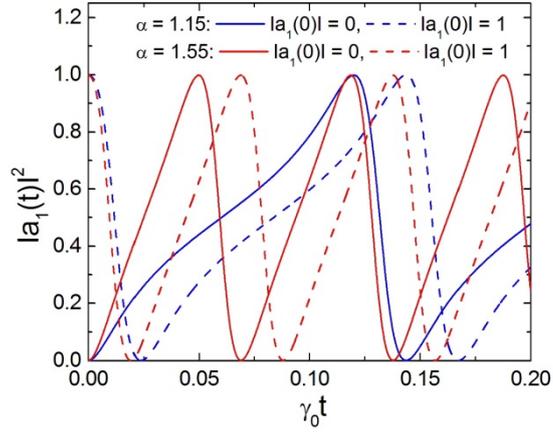

FIG. 3. (Color online) The probability of a QDE to be found in the excited state as a function of the normalized time in the strong pump regime for different values of parameter $\alpha > 1$ and initial conditions.

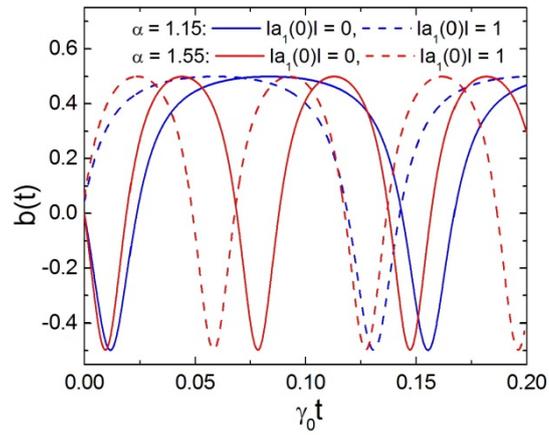

FIG. 4. (Color online) The dipole moment amplitude as a function of the normalized time in the strong pump regime for different values of parameter $\alpha > 1$ and initial conditions.

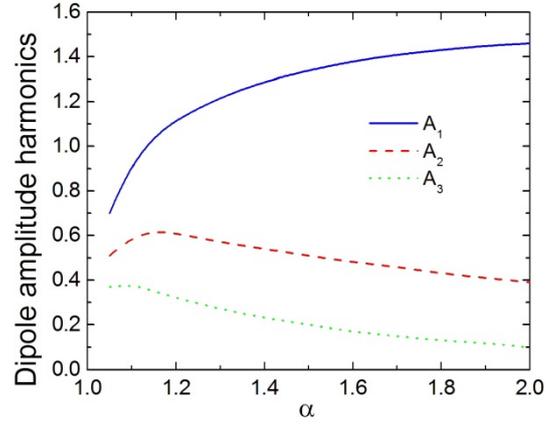

FIG. 5. (Color online) Dipole amplitude harmonics due to the anharmonicity in Rabi oscillations as a function of parameter $\alpha > 1$.

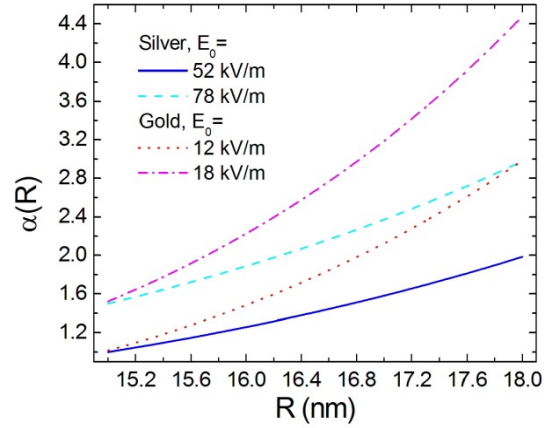

FIG. 6. (Color online) Dependences of parameter $\alpha$ on the distance between the QDE with the transition dipole moment $d_{10} = 4.8 \cdot 10^{-29} \, \text{C} \cdot \text{m}$ and the center of a 5-nm-radius MNP for different MNP materials and pump field strengths.